\documentclass[pra,aps,twocolumn]{revtex4}
\usepackage{amsmath,graphicx,epsfig}

\usepackage{euscript}
\usepackage{amsfonts}
\usepackage{amssymb}
\usepackage{float}

\def\prn#1{{\left(#1\right)}}

\def\sbrk#1{{\left[#1\right]}}
\def\abrk#1{{\langle#1\rangle}}
\def\ket#1{{|#1\rangle}}
\def\bra#1{{\langle#1|}}

\def\cg(#1,#2)(#3,#4)(#5,#6){\bra{#1,#2,#3,#4}#5,#6\rangle}

\def\ts#1{{_{\mbox{\scriptsize #1}}}}

\def\threej(#1,#2)(#3,#4)(#5,#6){\begin{pmatrix}#1&#3&#5\\#2&#4&#6\end{pmatrix}}
\def\sixj(#1,#2,#3)(#4,#5,#6){\begin{Bmatrix}#1&#2&#3\\#4&#5&#6\end{Bmatrix}}
\def\ninej(#1,#2,#3)(#4,#5,#6)(#7,#8,#9){\begin{Bmatrix}#1&#2&#3\\#4&#5&#6\\#7&#8&#9\end{Bmatrix}}

\def\sA{{\ensuremath{\EuScript A}}}

\def\sR{{\ensuremath{\EuScript R}}}
\def\sE{{\ensuremath{\EuScript E}}}

\def\mb{\mathbf}
\def\bs{\boldsymbol}

\newlength{\defbaselineskip}
\newcommand{\setlinespacing}[1]%
           {\setlength{\baselineskip}{#1 \defbaselineskip}}

\begin{document}

\title{A dual-isotope rubidium comagnetometer to search for anomalous long-range spin-mass (spin-gravity) couplings of the proton} 

\author{D. F. Jackson Kimball}
\email{derek.jacksonkimball@csueastbay.edu}
\affiliation{Department of Physics, California State University --
East Bay, Hayward, California 94542-3084, USA}

\author{I. Lacey}
\affiliation{Department of Physics, California State University --
East Bay, Hayward, California 94542-3084, USA}

\author{J. Valdez}
\affiliation{Department of Physics, California State University --
East Bay, Hayward, California 94542-3084, USA}

\author{J. Swiatlowski}
\affiliation{Department of Physics, California State University --
East Bay, Hayward, California 94542-3084, USA}

\author{C. Rios}
\affiliation{Department of Physics, California State University --
East Bay, Hayward, California 94542-3084, USA}

\author{R. Peregrina-Ramirez}
\affiliation{Department of Physics, California State University --
East Bay, Hayward, California 94542-3084, USA}

\author{C. Montcrieffe}
\affiliation{Department of Physics, California State University --
East Bay, Hayward, California 94542-3084, USA}

\author{J. Kremer}
\affiliation{Department of Physics, California State University --
East Bay, Hayward, California 94542-3084, USA}

\author{J. Dudley}
\affiliation{Department of Physics, California State University --
East Bay, Hayward, California 94542-3084, USA}

\author{C. Sanchez}
\affiliation{Department of Physics, California State University --
East Bay, Hayward, California 94542-3084, USA}

\date{\today}



\begin{abstract}
The experimental concept of a search for a long-range coupling between rubidium (Rb) nuclear spins and the mass of the Earth is described.  The experiment is based on simultaneous measurement of the spin precession frequencies for overlapping ensembles of $^{85}$Rb and $^{87}$Rb atoms contained within an evacuated, antirelaxation-coated vapor cell.  Rubidium atoms are spin-polarized in the presence of an applied magnetic field by synchronous optical pumping with circularly polarized laser light. Spin precession is probed by measuring optical rotation of far-off-resonant, linearly polarized laser light.  Simultaneous measurement of $^{85}$Rb and $^{87}$Rb spin precession frequencies enables suppression of magnetic-field-related systematic effects.  The nuclear structure of the Rb isotopes makes the experiment particularly sensitive to anomalous spin-dependent interactions of the proton.  Experimental sensitivity and a variety of systematic effects are discussed, and initial data are presented.
\end{abstract}



\maketitle

\section{Introduction}
\label{Sec:intro}

The connection between quantum theory and general relativity is one of the most important unsolved mysteries of modern physics, a mystery exacerbated by the dearth of experiments probing the rare intersections between these two theories.  One intersection between quantum effects and gravity that does offer potential for experimental tests is the question of how intrinsic spins interact with gravitational fields (see, for example, the review \cite{Ni10}).  According to general relativity, a purely tensor theory, the intrinsic spin of a particle is unaffected by the local gravitational field \cite{Kob62,Lei64,Heh90,Khr98,Sil05}.  However, in extensions of general relativity based on a Riemann-Cartan spacetime instead of a Riemann geometry, the gravitational interaction is described by a torsion tensor which can generate heretofore undetected spin-mass and spin-spin interactions \cite{Heh76,Sha02,Ham02,Kos08}.

In terms of quantum field theory phenomenology, the torsion tensor describes new scalar-pseudoscalar and vector-pseudovector gravitational interactions, corresponding, respectively, to spin-0 and spin-1 gravitons in addition to the usual spin-2 graviton associated with the tensor nature of standard gravity \cite{Nev80,Nev82,Car94}. New spin-1 and spin-0 partners of the usual spin-2 graviton also naturally arise in theoretical attempts to unify gravity and quantum mechanics, such as string theory and M-theory \cite{Fer77,Atw00,Sud02}, especially in the context of supersymmetry \cite{Gol86}.  It has recently been noted that a massless or nearly massless spin-0 component of gravity manifests as dark energy over cosmological distances \cite{Rat88,Wet88,Cal98,Fla09}.  The pseudoscalar component of such a field leads to an interaction that has the nonrelativistic form \cite{Fla09}:
\begin{align}
H_g = k \frac{\hbar}{c} \bs{\sigma}\cdot\bs{g}
\label{Eq:GDM-Hamiltonian}
\end{align}
where $k$ is a dimensionless parameter setting the scale of the new interaction, $\hbar$ is Planck's constant, $\bs{\sigma}$ is the intrinsic spin of the particle in units of $\hbar$, $\bs{g}$ is the Earth's gravitational field, and $c$ is the speed of light.  If the strength of the pseudoscalar coupling is the same as that of the tensor component of gravity, $k \approx 1$ \cite{Fla09}.

The Hamiltonian $H_g$ in Eq.~\eqref{Eq:GDM-Hamiltonian} manifestly violates the equivalence principle for intrinsic spins, offering a mechanism by which a gravitational field can be distinguished from an accelerating reference frame.  Furthermore, if $H_g \neq 0$, gravity would violate parity (P) and time-reversal (T) symmetries, and the spin-gravity coupling would be a source of additional CP-violation that might explain the observed matter-antimatter asymmetry of the universe \cite{Moh06}.  The possibility that gravity may violate discrete symmetries has led many authors over the past fifty years \cite{Lei64,Mor62,Kob63,Har76,Per78,Mac84,Muk99,Mas00,Pap02} to consider the possibility of such an interaction, which would imply the existence of a gravitational dipole moment (GDM) $k \hbar \bs{\sigma} / c$ for elementary particles.  One could envision a GDM as a separation between the center of inertial mass and the center of gravitational mass by a distance $k \hbar / m c$, where $m$ is the particle mass.  If the dimensionless coupling constant $k$ is of order unity, then the separation between the centers of inertial mass and gravitational mass is on the order of the Compton wavelength. Generally, theoretical models of gravity that accommodate such an interaction predict that $k \lesssim 1$  \cite{Lei64,Mor62,Kob63,Har76,Per78,Mac84,Muk99,Mas00,Pap02}.   However, to date, the most sensitive searches \cite{Ven92,Hec08} for such a spin-gravity coupling have set limits $k \lesssim 10$, still an order of magnitude away from the most theoretically interesting region of parameter space.  The central goal of the experiment described in the present paper is to probe spin-gravity interactions of the type described by Eq.~\eqref{Eq:GDM-Hamiltonian} at the $k \sim 1$ level.

\begin{figure*}
\center
\includegraphics[width = 5.5 in]{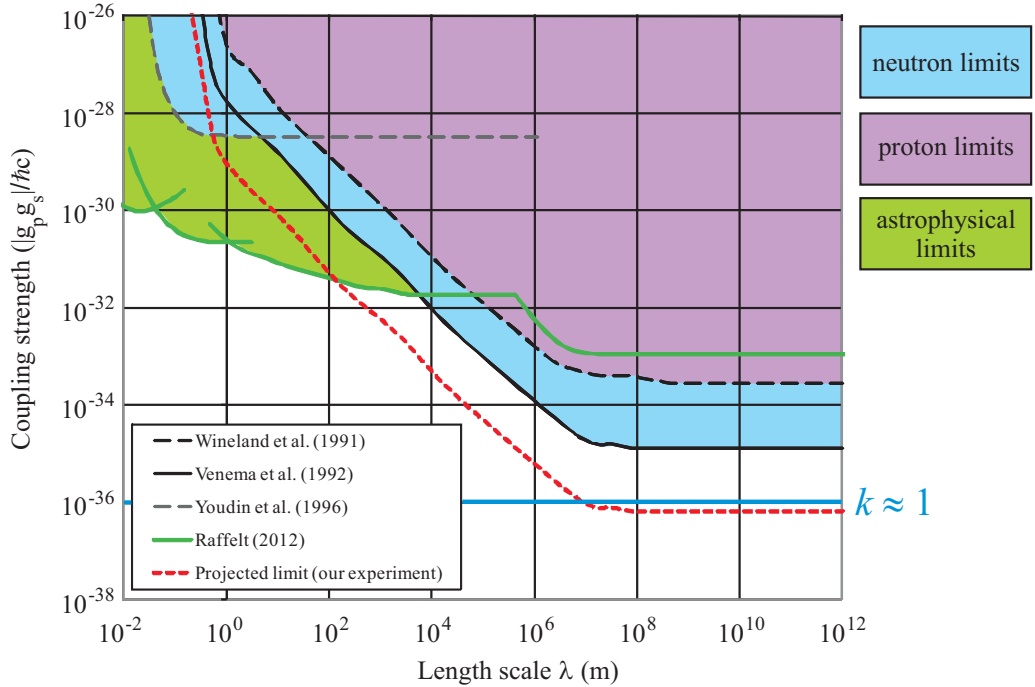}
\caption{\small{Existing experimental constraints (at the 2-$\sigma$ level) on nucleon monopole-dipole (scalar-pseudoscalar) couplings $\left| g_p g_s \right|/\hbar c$ as a function of the range $\lambda$ of the interaction ($g_p$ and $g_s$ are the pseudoscalar and scalar coupling constants, respectively).  Direct experimental constraints for the neutron are from Youdin {\it{et al.}} (1996) \cite{You96} at the laboratory-scale range and from Venema {\it{et al.}} (1992) \cite{Ven92} for the earth-scale range (excluded parameter space shaded blue); constraints for the proton are from the experiment of Wineland {\it{et al.}} (1991) \cite{Win91} (excluded parameter space shaded purple).  Astrophysical constraints for baryon couplings (excluded parameter space shaded green) are from the recent analysis of Raffelt (2012) \cite{Raf12}.  The nominal coupling strength for a scalar-tensor theory of gravity corresponding to $k \approx 1$ is represented by the blue line at the $\left| g_p g_s \right|/\hbar c \approx 10^{-36}$ level.  Constraints on monopole-dipole couplings of the electron spin to the mass of the earth (obtained by the University of Washington torsion pendulum experiment \cite{Hec08}, not shown) are similar to the constraints on the neutron spin from Ref.~\cite{Ven92}.  The potential sensitivity (dashed red line) of our proposed search could improve upon existing experimental limits on long-range ($\lambda \gtrsim 10^7~{\rm m}$) monopole-dipole couplings in general by an order of magnitude and for the proton spin in particular by three orders of magnitude.}}
\label{Fig:monopole-dipole-constraints-neutron-proton}
\end{figure*}

There are several experimental consequences of the existence of a GDM for an elementary particle.  The interaction described by Eq.~\eqref{Eq:GDM-Hamiltonian} leads to a gravity-induced splitting $\Delta E$ of the energy levels for spins oriented parallel and anti-parallel to $\bs{g}$:
\begin{align}
\Delta E = 2k \frac{\hbar g}{c} \approx 4k \times 10^{-23}~{\rm
eV}~. \label{Eq:GDM-EnergySplitting}
\end{align}
In addition, an interaction such as that described by Eq.~\eqref{Eq:GDM-Hamiltonian} generates a torque on spins immersed in a gravitational field, leading to spin precession about the axis of the local gravitational field with a frequency
\begin{align}
\frac{\Omega_g}{2\pi} = \frac{k g}{\pi c} \approx k \times 10^{-8}~{\rm Hz}~.
\label{Eq:GDM-PrecessionFrequency}
\end{align}
This spin precession not associated with the magnetic moments of the particles is the signature of the P- and T-violating spin-gravity coupling that is being searched for in our present experiment.

In principle, the effect of a spin-gravity coupling is indistinguishable from the interactions resulting from any heretofore undiscovered force-mediating pseudoscalar or vector particle.  In the literature, such new interactions are commonly parameterized using the Moody-Wilczek formalism for spin-0 particles \cite{Moo84}, recently extended by Dobrescu and Mocioiu to include spin-1 particles \cite{Dob06}.  In this context, the spin-gravity (or spin-mass) interaction searched for in our proposed experiment can be interpreted as a monopole-dipole coupling.  In general, the monopole-dipole coupling strength can be different for different elementary particles --- our proposed experiment is primarily sensitive to monopole-dipole couplings of the proton, whereas the previous best experimental limits are for neutron \cite{Ven92} and electron \cite{Hec08} couplings.  Consequently, our experiment has the potential to improve experimental constraints on monopole-dipole couplings of the proton by orders of magnitude compared to the best previous limit \cite{Win91}.

Figure~\ref{Fig:monopole-dipole-constraints-neutron-proton} presents a parameter exclusion plot showing existing direct experimental limits on monopole-dipole interactions of nucleons at various length scales, as well as astrophysical constraints inferred from the duration of the supernova SN 1987A neutrino burst in combination with data from searches for anomalous monopole-monopole forces \cite{Raf12}. (It should be noted that there are significant uncertainties related to dense nuclear matter effects in the analysis of the SN 1987A neutrino burst \cite{Jan96,Han01}.)  The blue horizontal line at the bottom of the plot shows the strength of the monopole-dipole interaction corresponding to $k \approx 1$.  The dashed red curve shows the projected sensitivity of our experiment.  Note that astrophysical limits for electron couplings, which are based on star cooling, are three orders of magnitude more restrictive than astrophysical limits on nucleon couplings.  Thus the astrophysical limits for electrons just reach the $k \approx 1$ regime, and are comparable to the projected sensitivity of our experiment.  Of interest are observations that the  white-dwarf luminosity function fits better with a small amount of anomalous energy loss at a level that would correspond to a pseudoscalar interaction with coupling strength corresponding to $k \approx 1$ \cite{Ise08}.  Furthermore, the observed period decrease of the pulsating white dwarf G117-B15A also favors some amount of extra cooling \cite{Ise10}.

In the present work, we describe a dual-isotope rubidium (Rb) comagnetometer well-suited for searching for a long-range monopole-dipole coupling between proton spins and the mass of the earth.  We show that the dual-isotope Rb comagnetometer as designed can achieve sufficient statistical sensitivity and rejection of known systematic errors so that present experimental limits on anomalous monopole-dipole couplings of the proton spin can be improved by orders of magnitude.  The basic concept of our experiment is to use synchronous laser optical pumping to generate transverse spin polarization of Rb atoms contained in an antirelaxation-coated cell \cite{Bal10}, and then employ off-resonant laser light to simultaneously measure the spin precession frequencies of $^{85}$Rb and $^{87}$Rb atoms in the presence of a magnetic field $\mb{B}$.  The ratio of the difference between the Rb precession frequencies divided by their sum,
\begin{align}
\sR = \frac{\Omega_{87}-\Omega_{85}}{\Omega_{87}+\Omega_{85}}~,
\end{align}
will be measured for a range of different magnetic fields.  Measurement of the ratio $\sR$ eliminates or reduces several common-mode sources of noise and systematic error.  Taking the difference $\Delta \sR$ between $\sR$ for $\mb{B}$ parallel with $\bs{g}$ and anti-parallel with $\bs{g}$ yields a signal proportional to the spin precession frequency caused by nonmagnetic interactions.  In this configuration, the valence electron spin of the Rb atoms effectively serves as an accurate comagnetometer for the Rb nuclear spins.

Our early work on this experiment is described in Ref.~\cite{Kim09}. In this early version of the experiment, a single linearly polarized laser beam, frequency-modulated at an integer multiple of the precession frequency, was used to measure nonlinear magneto-optical rotation (NMOR) induced by the spin precession of the Rb atoms \cite{Bud02-review,Ale05,Bud98,Bud00}.  While the NMOR measurements achieved our target statistical sensitivity to the Rb spin precession frequencies, a subtle systematic effect involving light shifts \cite{Kim09,Bud00-aoc} required us to modify our experimental approach.  The effect, known as alignment-to-orientation conversion, arises from the combined action of the magnetic field and optical electric field. Alignment-to-orientation conversion evolves spin polarization aligned along the linear polarization axis into spin polarization oriented along the light propagation (magnetic field) direction.  Atomic spins oriented along the light propagation direction generated ellipticity of the light field which in turn produced unacceptably large light-power- and magnetic-field-dependent shifts of the spin precession frequencies.  Our present experiment circumvents these systematic effects by temporally separating pump and probe stages, using unmodulated probe light, and by a choice of experimental geometry where the probe beam propagates in a direction orthogonal to the magnetic field (Sec.~\ref{Sec:systematic-effects}).

\section{Experimental concept}
\label{Sec:concept}

As noted in Sec.~\ref{Sec:intro}, a long-range spin-mass or spin-gravity interaction can be parameterized in terms of a GDM $\bs{\kappa}$ via $\bs{\kappa} = \chi \bs{\sigma}$, where $\chi = k\hbar/c$ is the ``gyro-gravitational ratio'' for the particle.  The $\chi$ for Rb atoms can be calculated in terms of $\chi_e$ and $\chi_p$, the gyro-gravitational ratios for the electron and proton, respectively, using the shell model to describe the nuclei.  In a given ground-state hyperfine level with total angular momentum $F$, the atomic GDM $\bs{\kappa}\ts{atom}(F)$ is
\begin{align}
\bs{\kappa}\ts{atom}(F) &= \chi\ts{atom}(F) \bs{F} \\
&= \frac{\abrk{ \bs{S}_e \cdot \bs{F} }}{F(F+1)}\abrk{\bs{F}} \chi_e + \frac{\abrk{ \bs{I} \cdot \bs{F} }}{F(F+1)}\abrk{\bs{F}} \chi\ts{nucl}~, \nonumber
\end{align}
where $\bs{S}_e$ is the electron spin, $\bs{I}$ is the nuclear spin, and $\chi\ts{nucl}$ is the nuclear gyro-gravitational ratio.  In the nuclear shell model \cite{Kli52}, both Rb isotopes have valence protons, and so the nuclear spin, magnetic moment, and nuclear GDM are, to a good approximation, due entirely to the proton.  The nuclear GDM $\bs{\kappa}\ts{nucl}$ is thus given by
\begin{align}
\bs{\kappa}\ts{nucl} = \chi\ts{nucl} \bs{I} = \frac{\abrk{ \bs{S}_p \cdot \bs{I} }}{I(I+1)}\abrk{\bs{I}} \chi_p~,
\end{align}
where $\bs{S}_p$ is the proton spin and we have assumed, as do most theoretical models \cite{Lei64,Fla09,Mor62,Kob63,Har76,Per78,Mac84,Muk99,Mas00,Pap02}, that there is no contribution from orbital angular momentum.  Relevant parameters based on the above considerations are presented for the two Rb isotopes in the probed ground-state hyperfine levels in Table~\ref{Table:RbGDMparameters}.

\begin{table*}
\caption{Parameters determining gyro-gravitational ratios $\chi$ and gyromagnetic ratios $\gamma = g_F\mu_0$ for $^{85}$Rb and $^{87}$Rb in the ground-state hyperfine levels of interest.}
\medskip \begin{tabular}{lcccccc} \hline \hline
\rule{0ex}{3.6ex} Atom~~ & ~~Ground electronic state~~ & ~~Nuclear spin~~ & ~~Total angular momentum~~ & ~~$g$-factor~~ & ~~Proton state~~ &  ~~$\chi\ts{atom}$ \\
\hline
\rule{0ex}{3.6ex} $^{85}$Rb & $5s~^{2}S_{1/2}$ & $I=5/2$ & $F=3$ & $g_F = 1/3$ & $4f_{5/2}$ & $\frac{1}{6}\chi_e - \frac{5}{42} \chi_p$  \\
\rule{0ex}{3.6ex} $^{87}$Rb & $5s~^{2}S_{1/2}$ & $I=3/2$ & $F=2$ & $g_F = 1/2$ & $3p_{3/2}$ & $\frac{1}{4}\chi_e + \frac{1}{4} \chi_p$  \\
~ & ~ & ~ & ~ & ~ & ~ &  ~ \\
\hline \hline
\end{tabular}
\label{Table:RbGDMparameters}
\end{table*}

Ignoring temporarily other causes of spin precession, the spin-precession frequencies for $^{85}$Rb and $^{87}$Rb in the presence of the magnetic field $\mb{B}$ and Earth's gravitational field $\bs{g}$ are
\begin{align}
\Omega_{85} & \approx \left| \gamma_{85}B + \prn{\frac{1}{6}\chi_e - \frac{5}{42} \chi_p} g\cos\phi  \right|~, \\
\Omega_{87} & \approx \left| \gamma_{87}B + \prn{\frac{1}{4}\chi_e + \frac{1}{4} \chi_p} g\cos\phi  \right|~,
\end{align}
where $\phi$ is the angle between $\mb{B}$ and $\bs{g}$, $\gamma_{85}$ and $\gamma_{87}$ are the gyromagnetic ratios ($\gamma = g_F\mu_0$, where $g_F$ is the Land\'e $g$-factor and $\mu_0$ is the Bohr magneton) and the light propagation direction is along $\mb{B}$.  In the above we neglect contributions to the spin-precession frequency second-order in $g$. To analyze the data, we construct the following ratio:
\begin{align}
\sR = \frac{\Omega_{87} - \Omega_{85}}{\Omega_{87}+\Omega_{85}}~.
\label{Eq:SpinPrecessionRatio}
\end{align}
To first order assuming $\gamma B \gg \chi_e g,\chi_p g$ and neglecting the effects of the nuclear magnetic moments, we have
\begin{align}
\sR_\pm \approx \prn{\frac{\gamma_{87}-\gamma_{85}}{\gamma_{87}+\gamma_{85}}} \prn{1 \pm 2.06 \frac{\chi_p g \cos\phi}{\mu_0 B}}~,
\label{Eq:R}
\end{align}
where $\sR_+$ is for positive $B$ and $\sR_-$ is for negative $B$ (relative to $\bs{g}$).  There is first-order cancelation of the effects of an electron GDM in the ratio $\sR$, and near unity sensitivity to the effects of a proton GDM.  Measuring $\Delta \sR = \sR_+ - \sR_-$ yields a signal proportional only to the proton GDM:
\begin{align}
\Delta \sR \approx 4.12 \prn{\frac{\gamma_{87}-\gamma_{85}}{\gamma_{87}+\gamma_{85}}} \prn{\frac{\chi_p g \cos\phi}{\mu_0 B}}~.
\label{Eq:DeltaR}
\end{align}
The first-order cancelation of the electron GDM contribution to $\Delta \sR$ is a result of the fact that we measure spin precession in the $F = I+1/2$ ground state hyperfine level for both isotopes, so electron couplings contribute in nearly identical ways to the measured values of $\Omega_{85}$ and $\Omega_{87}$.

\section{Experimental setup}
\label{Sec:setup}

\begin{figure*}
\center
\includegraphics[width = 5.5 in]{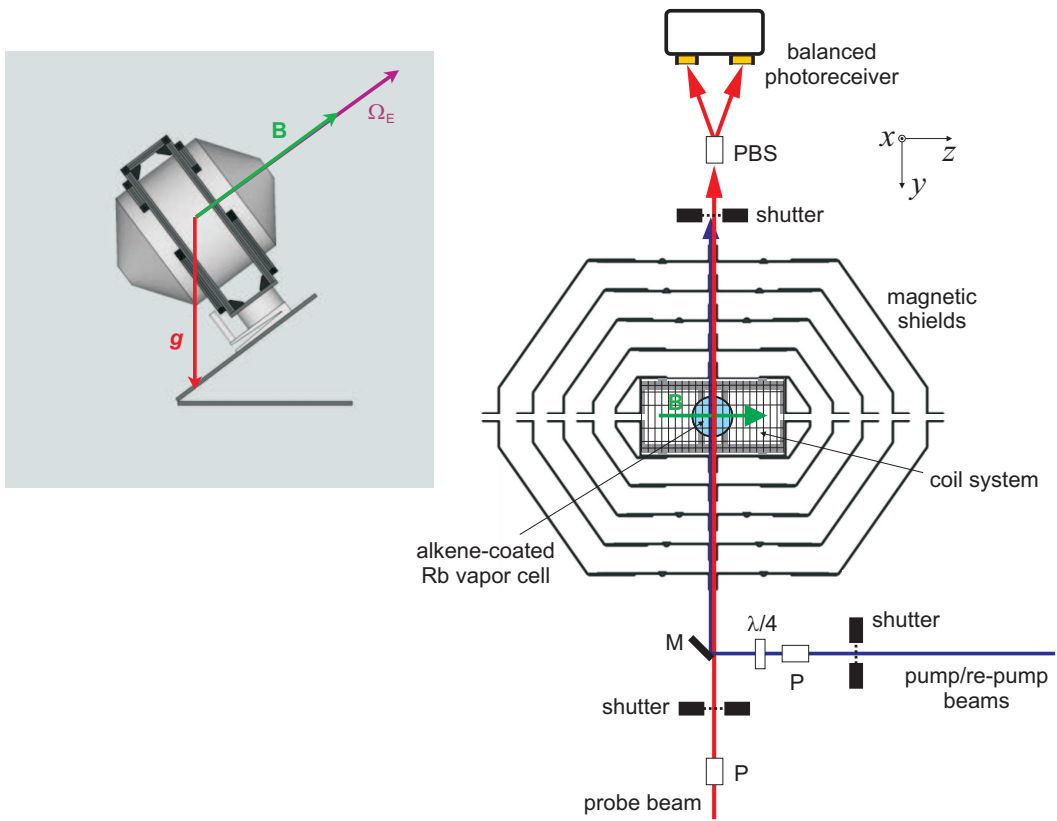}
\caption{\small{Right-hand side: schematic diagram of the experimental setup used to measure  the spin precession frequencies of $^{85}$Rb and $^{87}$Rb and to search for a long-range spin-mass (spin-gravity) coupling (P = linear polarizer, M = mirror, PBS = polarizing beamsplitter, $\lambda/4$ = quarter wave plate).  Picture, upper left: schematic of the apparatus used to align the magnetic shield system and magnetic field $\mb{B}$ along the Earth's rotation axis $\Omega_E$, and geometrical relationship to the local gravitational field $\bs{g}$ (this geometry is chosen to control systematic errors related to the earth's rotation, see Sec.~\ref{Sec:gyro-compass}).}}
\label{Fig:expt-setup}
\end{figure*}

A schematic diagram of the experimental setup used to carry out simultaneous measurement of $\Omega_{85}$ and $\Omega_{87}$ is shown in Fig.~\ref{Fig:expt-setup}. At the heart of the experiment is a natural isotopic mixture of Rb vapor (72.2\% $^{85}$Rb, 27.8\% $^{87}$Rb) contained within an evacuated (residual pressure $\approx 10^{-6}~{\rm torr}$) spherical alkene-coated glass cell (diameter = 5~cm).  The alkene coating is 1-nonadecene [$\rm{ CH_2-CH(CH_2)_{16}-CH_2 }$] and the cell was prepared according to procedures outlined in Ref.~\cite{Bal10}.  The particular cell we are using was measured to have longitudinal spin relaxation times $T_1 \approx 5~{\rm s}$ limited by exchange of atoms between the spherical bulb of the cell and the stem which contains the Rb reservoir.  Under typical operating conditions, the spin relaxation rate due to wall collisions is significantly smaller than the relaxation rate due to spin-exchange collisions between the Rb atoms.

The vapor cell is mounted inside a frame manufactured of HDPE (High Density Polyethylene) which is fit inside the innermost layer of a five-layer magnetic shield (manufactured by Amuneal Inc.) made of a 1-mm thick high-permeability alloy, annealed in a hydrogen atmosphere.  Each layer of the shield consists of a cylindrical center piece and two removable end caps.  The outer layers of the shield are spaced by styrofoam (polymerized in place) and the innermost layer is spaced by melamine foam to reduce acoustic noise.  Four ports for access to the inside of the shields are available on the cylindrical pieces and one port is available on each end cap.  The shielding factor of the entire five-layer magnetic shield system was measured for a nearly identical design to be better than $10^7$ \cite{Xu06}.  The foam spacing between the shield layers provides thermal insulation in addition to mechanical support.  The temperature of the innermost shield layer is stabilized at $30^\circ{\rm C}$ by a J-KEM Model 210 temperature controller using a T-type thermocouple attached to the inner surface of the shield layer for temperature measurement and resistive heating with a twisted pair of wires wrapped about the outside of the innermost shield layer.  Stabilizing the shield temperature serves two functions: (1) it reduces temperature-related drifts of residual magnetic fields from the innermost shield and (2) it provides a stable, elevated temperature environment for the Rb cell yielding vapor densities of $\approx 2 \times 10^{10}~{\rm atoms/cm^3}$.

A system of nine separate coils are wound in grooves cut into the frame mounted inside the innermost layer of the shield.  The system of coils was designed to provide, over the volume of the Rb vapor cell, uniform magnetic fields in three orthogonal directions ($B_x$, $B_y$, and $B_z$), linear magnetic field gradients in five directions ($dB_x/dx$, $dB_z/dz$, $dB_x/dz$, $dB_y/dz$, $dB_y/dx$), and a quadratic gradient along the shield axis ($d^2B_z/dz^2$).  As a consequence of Maxwell's equations ($\nabla \cdot \mb{B} = 0$ and $\nabla \times \mb{B} = 0$), control over the five linear magnetic field gradients is sufficient to provide compensation of all nine possible linear gradients. Based on computer modeling (using the Amperes program from Integrated Engineering Software Inc.), the uniformity of the magnetic fields and linearity/quadracity of the field gradients generated by the coil system is at a part per thousand over the cell volume for typical applied currents.  It should be noted that effects of uncompensated magnetic-field gradients are significantly reduced by motional averaging \cite{Pus06-grad} (effects are quadratic in the the magnitude of the gradient). The coils are in series with a set of ultra-stable, low temperature coefficient (low TC) resistors (Caddock Type USF 200 Series, zero nominal TC with TC $\lesssim 2~{\rm ppm/K}$).  The voltage for the $B_z$ coil is supplied by a precision DC voltage source (Krohn-Hite Model 523 calibrator, stability $\pm 1~{\rm ppm}$) and voltages for the coils controlling $B_x$, $B_y$, and field gradients are computer generated with a digital-to-analog-converter (DAC, National Instruments PCI-6733).

In order to measure $\Omega_{85}$ and $\Omega_{87}$, a system of shutters (Stanford Research SR474) is used to implement a temporally separated pump/probe measurement scheme. During the optical pumping stage (duration = 1~s), Rb atoms are illuminated by two collinear, circularly polarized pump beams propagating along $-\mb{\hat{y}}$ (orthogonal to $\mb{B}$ which is along $z$), one tuned to the center of the Doppler-broadened $^{85}$Rb D2 $F=3 \rightarrow F'$ resonance and the other tuned to the center of the Doppler-broadened $^{87}$Rb D1 $F=2 \rightarrow F'=1$ resonance ($F$, $F'$ are the total atomic angular momenta of the ground and excited states, respectively).  The 780-nm D2 pump beam is generated by a distributed feedback laser diode (EYP-DFB-0780-00080-1500-TOC03 from Eagleyard Photonics) and the 795-nm D1 pump beam is produced by a tunable external-cavity diode laser (Toptica DL100). The pump beams are amplitude-modulated at frequencies close to the respective Larmor frequencies of the isotopes using electro-optic modulators (EOMs, ThorLABs E0-AM-NR-C1, not shown in Fig.~\ref{Fig:expt-setup}) placed between crossed calcite linear polarizers.  The duty cycle for both pump beams is 20\%; during the period when the EOMs transmit the pump light, the power of the D2 pump beam incident on the Rb atoms is $\approx 55~\mu{\rm W}$ and the power of the D1 pump beam is $\approx 150~\mu{\rm W}$.  These parameters were chosen to maximize the transverse spin polarization for both Rb isotopes. This synchronous optical pumping generates atomic spin polarization transverse to $\mb{B}$ in both isotopes precessing at their respective Larmor frequencies.  A third, collinear, circularly polarized re-pump beam tuned to the center of the Doppler-broadened $^{87}$Rb D2 $F=1 \rightarrow F'$ resonance transfers $^{87}$Rb atoms pumped into the unobserved $F=1$ ground-state hyperfine level back into the $F=2$ hyperfine level (which, taking into account natural isotopic abundances, yields approximately equal signals for both isotopes).  The 780-nm re-pump beam is produced by tunable external-cavity diode laser (New Focus TLM 7000), with power $\approx 750~\mu{\rm W}$. The diameters of the pump and re-pump beams are $\approx 2~{\rm mm}$.

During the optical probing stage (duration = 1~s), a shutter blocks the pump and re-pump beams, and shutters open to allow a linearly polarized probe laser beam to propagate along $-\mb{\hat{y}}$ through the vapor and into a polarimeter.  The 780-nm D2 probe beam is produced by another tunable external-cavity diode laser (Toptica DL100).  The $^{85}$Rb and $^{87}$Rb precession frequencies, $\Omega_{85}$ and $\Omega_{87}$, are measured by observing optical rotation of the probe light.  The frequency of the probe beam is tuned $\approx 3~{\rm GHz}$ below the center frequency of the Doppler-broadened $^{87}$Rb D2 $F=2 \rightarrow F'$ resonance, the power is $\approx 200~\mu{\rm W}$, and the beam diameter is $\approx 2~{\rm mm}$.    Prior to entering the vapor cell, the probe beam passes through an antireflection-coated Glan Thomson linear polarizer (calcite, extinction ratio $5 \times 10^5 : 1$).  After exiting the vapor cell, the beam is analyzed by a polarimeter consisting of a Wollaston prism polarizing beamsplitter (calcite, extinction ratio $10^5 : 1$) whose output rays are detected with a balanced photoreceiver (New Focus Model 2307). The signal from the photoreceiver is sent to a preamplifier (Stanford Research Systems SR560) and then recorded on computer using an analog-to-digital converter (National Instruments PCIe-6361) using a routine written in LabVIEW.  The time base for the data acquisition is provided by a 10~MHz signal from a Rb atomic frequency standard (Stanford Research Systems SIM940, short-term stability $\lesssim 2 \times 10^{-12}$ in 100~s) that is GPS-disciplined with a 1 PPS signal (from a Communication Navigation Surveillance Inc. CNS Clock II, with a long-term accuracy better than a part in $\approx 10^{12}$).  The accurate time base ensures that $\Omega_{85}$ and $\Omega_{87}$ can be measured at the $10^{-8}~{\rm Hz}$ level over a long period of time for data averaging.  The pump and re-pump lasers are frequency stabilized using dichroic atomic vapor laser locks (DAVLLs) \cite{Cor98,Yas00}.  The probe, pump, and re-pump beam spectral purities are monitored with Fabry-Perot interferometers (ThorLABs SA200-5B) and the light powers of each beam transmitted through separate uncoated Rb reference cells (natural isotopic mixture) are monitored to ensure that the lasers remain properly tuned. (The laser frequency locking and diagnostics setups are not shown in Fig.~\ref{Fig:expt-setup}.)

The picture in the upper left corner of the experimental setup diagram (Fig.~\ref{Fig:expt-setup}) depicts the magnetic shield mount used for mechanical alignment of the shield axis $z$ along the Earth's rotation axis $\bs{\hat{\Omega}_E}$, which is important for control of a systematic error related to the Earth's rotation (Sec.~\ref{Sec:gyro-compass}).  The outermost shield layer is held in place with an aluminum frame, which is bolted to precision tilt and rotation stages (Newport TGN160 and UTR120, respectively) attached to an optical breadboard tilted from horizontal by an angle approximately equal to the latitude of the laboratory ($37^\circ~39'~24''$~N).  By surveying the laboratory (using Google Earth as well as GPS signals) and using an alignment laser propagating along the shield axis ($z$) with a path length of $\approx 3~{\rm m}$, we are able to mechanically align the shield axis with the Earth's rotation axis to within $0.3^\circ$.  Prior to measurement of $\Omega_{85}$ and $\Omega_{87}$, $\mb{B}$ is carefully aligned along the $z$-axis using a laser beam split off from the 780-nm probe beam by measuring NMOR with frequency modulated light \cite{Kim09,Bud02-review,Bud02,Yas03}.  The accuracy of the alignment of $\mb{B}$ along the light propagation direction $\mb{k}$ using NMOR \cite{Pus06-tilt} is much greater than the mechanical alignment accuracy of the shield axis, so the alignment of $\mb{B}$ parallel with $\bs{\hat{\Omega}_E}$ is achieved with an uncertainty of $\approx 0.3^\circ$.

\section{Initial data and projected statistical sensitivity}
\label{Sec:data-sensitivity}

\begin{figure}
\includegraphics[width = 3.25 in]{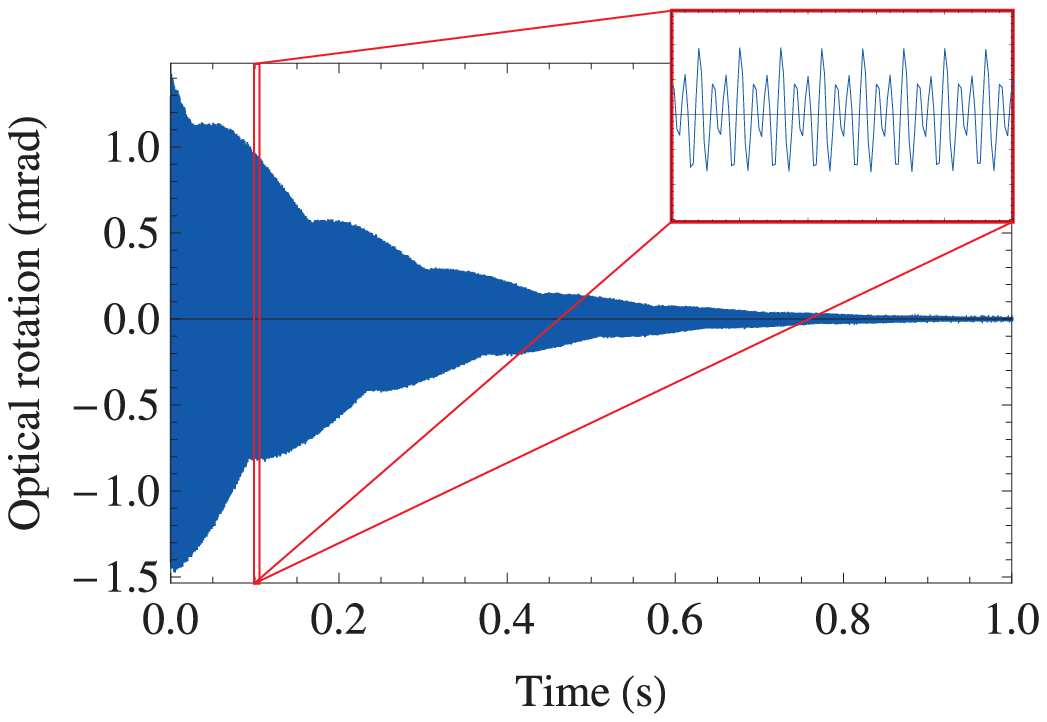}
\includegraphics[width = 3.25 in]{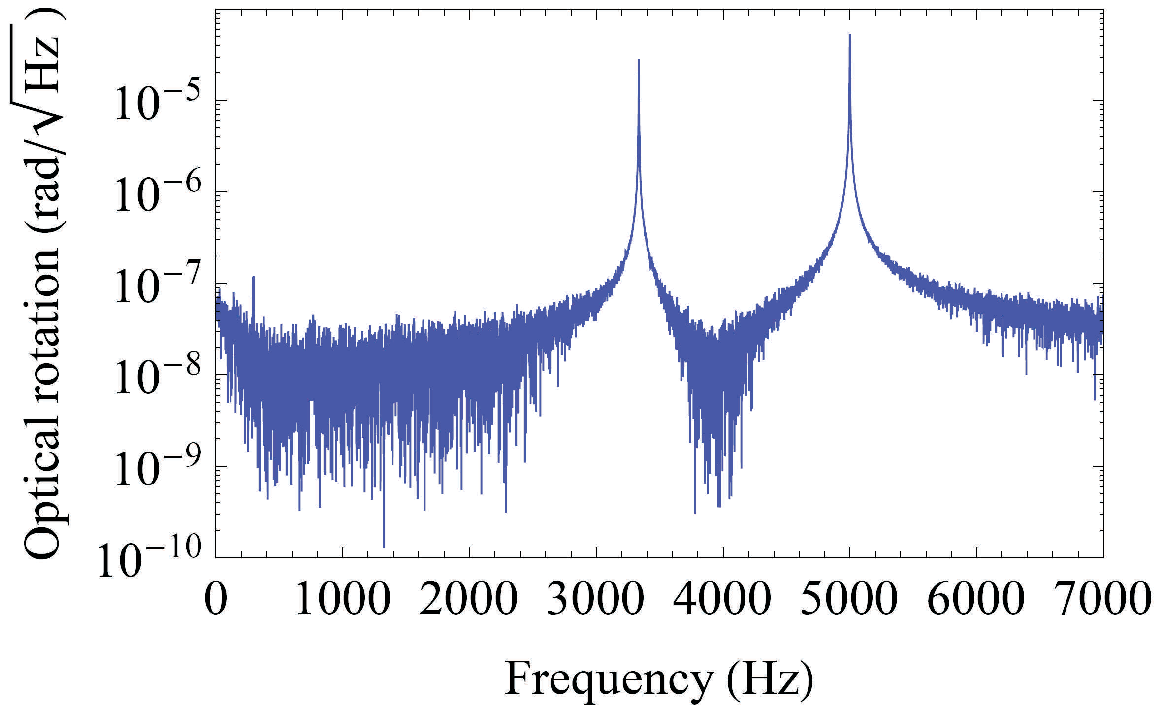}
\caption{\small{Upper plot: sample time-domain data from pump/probe measurement of Rb spin precession using optical rotation.  Inset in upper right corner shows a subset of the data of total duration 10~ms (highlighted by the red box on the complete data set).  In the inset, fast beating is observed between sinusoidal signals with frequencies $\Omega_{85}$ and $\Omega_{87}$.  The applied magnetic field corresponds to $|\mb{B}| \approx 7.1443~{\rm mG}$. Slow beating can be observed in the main time-domain plot between a dominant signal from $^{85}$Rb atoms in the $F=3$ ground state hyperfine level and a smaller-amplitude signal from $^{85}$Rb atoms in the $F=2$ ground state hyperfine level (the absolute value of the Land\'e $g$-factors of the two ground state hyperfine levels differ due to the nuclear magnetic moments, see Sec.~\ref{Sec:nuclear-magnetic-moments}). Lower plot: Fourier transform of the time-domain data, showing resonances at $\Omega_{85}$ (lower-frequency peak) and $\Omega_{87}$ (higher-frequency peak).}}
\label{Fig:GDM-data}
\end{figure}

Sample data acquired during the probe sequence are shown in Fig.~\ref{Fig:GDM-data}.  The upper plot shows the optical rotation signal acquired in the time domain for an applied field of $B \approx 7.1443~{\rm mG}$ oriented in the $\mb{\hat{z}}$ direction, and the lower plot shows the absolute value of the Fourier transform of the data set (carried out using a data analysis routine written in Mathematica).  In the frequency domain, distinct resonant peaks in the Fourier transform can be identified and correspond to $\Omega_{85}$ and $\Omega_{87}$.  The dominant contribution to these signals is from atoms in the resonantly pumped $^{85}$Rb $F=3$ ground state and the $^{87}$Rb $F=2$ ground state, since the pump and re-pump laser beam parameters are optimized for transverse spin polarization of these states and the probe beam is tuned closest to optical resonance with transitions from these states.

In the time domain a slow beating is observed in the optical rotation signal. The additional frequency component responsible for the beating arises from spin precession of $^{85}$Rb atoms in the $F=2$ ground state, which has a slightly different Land\'e $g$-factor magnitude than the $F=3$ ground state because of the nuclear magnetic moment (this effect is discussed in detail in Sec.~\ref{Sec:nuclear-magnetic-moments}).  The $^{85}$Rb $F=2$ ground state is slightly polarized by nearly synchronous re-population pumping: a fraction of the $^{85}$Rb atoms optically pumped from the $F=3$ ground state spontaneously decay from the ${\rm 5~^2P_{3/2}}$ excited state back to the $F=2$ ground state; the absolute values of the Land\'e $g$-factors are close enough in value that for sufficiently small magnetic fields a detectable transverse spin polarization in the $^{85}$Rb $F=2$ ground state can be created during the pumping stage.  ($^{87}$Rb atoms in the $F=1$ ground state contribute a much smaller amplitude signal for a variety of reasons, discussed in Sec.~\ref{Sec:nuclear-magnetic-moments}.)

Data analysis is carried out by fitting subsets of the Fourier transformed optical rotation signal centered around $\Omega_{85}$ and $\Omega_{87}$ to a Lorentzian function $S(\omega)$:
\begin{align}
S(\omega) = \sqrt{ \sbrk{ \frac{\alpha}{ 1 + \prn{ \frac{\omega-\Omega}{\Gamma} }^2 } }^2 + \sbrk{ \frac{\beta (\omega-\Omega)/\Gamma}{ 1 + \prn{ \frac{\omega-\Omega}{\Gamma} }^2 } }^2 }~,
\end{align}
where $\alpha$ and $\beta$ are the amplitudes of the imaginary and real components of the signal, respectively, $\omega$ is the frequency, $\Omega$ is the resonant spin-precession frequency, and $\Gamma$ is the resonance width (corresponding to the spin relaxation rate).  To account for spin precession of $^{85}$Rb atoms in the $F=2$ ground state, additional real and imaginary Lorentzian components of the signal can be included in the fitting function for $\Omega_{85}$.  Under typical operating conditions, $\Gamma/(2\pi) \approx 1~{\rm Hz}$.

Fits to the data demonstrate a statistical sensitivity to the spin precession frequency $\Omega$ of
\begin{align}
\frac{\delta \Omega}{2\pi} \approx 100~{\rm \mu Hz}
\label{Eq:GDM-measured-precision}
\end{align}
for a 1~s measurement.  Ultimately, the shot-noise-limited (SNL) sensitivity $\delta \Omega\ts{SNL}$ of a spin-polarized atomic sample to precession frequencies is determined by the total number of atoms $N$ and the relaxation rate $\Gamma\ts{rel}$ of the atomic spin polarization (for measurement times $\tau \gg \Gamma\ts{rel}^{-1}$ \cite{Auz06,Bud07,Bud13}):
\begin{align}
\delta \Omega\ts{SNL} \approx \sqrt{ \frac{\Gamma\ts{rel}}{N \tau} }~.
\label{Eq:FundamentalSensitivity-atoms}
\end{align}
Under our experimental conditions, $N \approx 10^{12}~{\rm atoms}$ and $\Gamma\ts{rel} \approx 2\pi \times 1~{\rm Hz}$ (limited by spin-exchange collisions), yielding $\delta\Omega\ts{SNL} \approx 2\pi \times 0.4~{\rm \mu Hz}$ for $\tau = 1~{\rm s}$.  This suggests that the measurement is not presently shot-noise-limited, and further reduction in technical noise would permit even better statistical sensitivity.

We are in the process of making several modifications to the apparatus in order to improve the sensitivity.  Noise from the balanced photoreceiver presently exceeds the photon-shot-noise limit by a factor of $\approx 5$, and therefore we are upgrading the balanced photoreceiver in order to achieve photon-shot-noise-limited polarimetry.  Optimization of the probe light power and detuning should enable further improvement in sensitivity, since the photon shot-noise limit presently exceeds the atomic shot-noise limit by over an order of magnitude.

Although comagnetometry significantly reduces magnetic-field-related noise and systematic effects from acquisition-to-acquisition, in our configuration it does not improve the statistical uncertainty for a single acquisition. Therefore magnetic field noise can degrade the sensitivity.  In particular, magnetic field noise due to thermal Johnson currents in the innermost mu-metal shield is estimated to contribute noise at the $100~{\rm \mu Hz}/\sqrt{\rm Hz}$ level \cite{Nen96,Lam99,Kom03}.  We are in the process of replacing the innermost mu-metal shield with a non-conducting ferrite shield, which has been demonstrated to reduce thermal magnetic field noise \cite{Kor07}.

We expect that these improvements to our apparatus should enable sensitivity to atomic spin precession at the $10~{\rm \mu Hz}$ level for a 1~s measurement.  Collecting data for $\approx 10^6~\rm{s}$ would then yield a statistical sensitivity of $\approx 10^{-8}$~Hz to anomalous spin-precession, sufficient to search for a GDM signal corresponding to $k \sim 1$ (Eq.~\ref{Eq:GDM-PrecessionFrequency}).

\section{Systematic effects}
\label{Sec:systematic-effects}

\subsection{General considerations}
\label{Sec:systematic-effects-general}

For a general consideration of systematic errors, it is helpful to characterize the ways in which additional contributions to the spin-precession frequencies for $^{85}$Rb and $^{87}$Rb beyond Larmor precession can enter the expressions for $\Omega_{85}$ and $\Omega_{87}$, and, crucially, the comagnetometer signal $\Delta \sR$ from which we will extract the GDM coupling.

One useful way to characterize systematic errors is to separate contributions to the spin precession frequencies into those that reverse sign when the direction of $\mb{B}$ is changed relative to $\bs{g}$ ($B$-odd terms, $\Omega_o$) and those that do not reverse sign ($B$-even terms, $\Omega_e$).  Depending on the orientation of $\mb{B}$ relative to $\bs{g}$, for each isotope we obtain two different precession frequencies
\begin{align}
\Omega_{\pm} = \Omega_L + \Omega_e \pm \Omega_o~,
\end{align}
where $\Omega_L$ is the appropriate Larmor frequency.  Somewhat counter-intuitively, any effect that causes spin precession in a fixed sense contributes a $B$-odd term.  This is because reversal of $\mb{B}$ reverses the sense of Larmor precession, and it is the absolute value of the spin precession frequency that is measured in the experiment.  Thus spin-precession due to a GDM coupling is a $B$-odd term contributing to $\Omega_o$, and consequently $B$-odd systematic errors are not suppressed in the comagnetometer signal $\Delta \sR$. Assuming $\Omega_e = 0$ and a $B$-odd systematic effect adding to the $^{87}$Rb precession frequency, we have
\begin{align}
\Delta \sR \approx 9.6 \prn{\frac{\gamma_{87}-\gamma_{85}}{\gamma_{87}+\gamma_{85}}} \frac{\Omega_o}{\mu_0 B}~.
\label{Eq:B-odd-shift}
\end{align}
Therefore $B$-odd systematic effects must be suppressed or accounted for by other means.

In the case of perfect magnetic-field reversal, there is no contribution of $B$-even terms to $\Delta \sR$.  However, if magnetic-field reversal is imperfect by an amount $\delta B$, $B$-even systematic effects can lead to a nonzero $\Delta \sR$.  Assuming $\Omega_o=0$ and a $B$-even systematic effect adding to the $^{87}$Rb precession frequency, we have
\begin{align}
\Delta \sR \approx 4.8 \prn{\frac{\gamma_{87}-\gamma_{85}}{\gamma_{87}+\gamma_{85}}} \frac{\Omega_e}{\mu_0 B} \frac{\delta B}{B}~.
\label{Eq:B-even-shift}
\end{align}
Thus $B$-even systematic effects are suppressed relative to $B$-odd effects by a factor $\sim \delta B/B$ which in our experiment can be made $\lesssim 10^{-9}$ by taking advantage of the high sensitivity of the setup to magnetic fields (under typical experimental conditions, the applied magnetic field is $\sim 10~{\rm mG}$ and the magnetometric sensitivity of the apparatus is $\delta B \sim 10^{-11}~{\rm G/\sqrt{Hz}}$).

Because our experiment employs a scalar measurement scheme (see, e.g., Refs.~\cite{Bud07,Bud13}) where the dominant contribution to the spin precession frequency is from Larmor precession induced by the magnetic field $\mb{B}$, we can also characterize spin precession in terms of contributions $\Omega_{||}$ that add linearly to $\Omega_L$ and contributions $\Omega_\bot$ that add in quadrature to Larmor precession:
\begin{align}
\Omega &= \sqrt{ \prn{\Omega_L + \Omega_{||} }^2 + \prn{\Omega_{\bot}}^2 }~,\\
&\approx  \Omega_L + \Omega_{||} + \frac{\Omega_{\bot}^2}{2\Omega_L}~,
\end{align}
where we have assumed that $\Omega_{||},\Omega_{\bot} \ll \Omega_L$.

Certain systematic errors (such as light shifts) are suppressed by arranging the experimental geometry so that they contribute to $\Omega$ primarily as $\Omega_{\bot}$.  Of course, imperfections in alignment inevitably mean that there is some contribution of such systematic errors to both $\Omega_{||}$ and $\Omega_\bot$:
\begin{align}
\Omega_{||} &= \Omega\ts{err} \sin \varphi \approx \varphi \Omega\ts{err}~, \\
\Omega_\bot &= \Omega\ts{err} \cos \varphi \approx \prn{ 1 - \frac{\varphi^2}{2} } \Omega\ts{err}~, \\
\Omega &\approx \Omega_L + \varphi \Omega\ts{err} + \frac{\Omega\ts{err}^2}{2\Omega_L}~,
\label{Eq:orthogonal-shift}
\end{align}
where $\Omega\ts{err}$ is the amplitude of the systematic error and $\varphi$ is the misalignment angle from perfect orthogonality to the leading field contribution.  Experimentally, mechanical alignment of the system can in most cases achieve at best $\varphi \lesssim 5 \times 10^{-3}~{\rm rad}$ ($0.3^\circ$).  However, in the case of the alignment of $\mb{B}$ parallel with or orthogonal to the light propagation direction (represented by the wave vector $\mb{k}$), much better results can be achieved by employing nonlinear magneto-optical effects that depend on the angle between $\mb{B}$ and $\mb{k}$ \cite{Pus06-tilt}: $\varphi \lesssim 10^{-5}~{\rm rad}$ can be achieved under typical operating conditions.

Other systematic errors cannot be sufficiently suppressed using the above experimental geometry (for example, the gyroscopic error introduced by rotation of the Earth \cite{Ven92,Hec08,Smi11}).  In these cases, the experiment is arranged so that the error contributes to $\Omega$ primarily as $\Omega_{||}$:
\begin{align}
\Omega_{||} &= \Omega\ts{err} \cos \varphi \approx \prn{ 1 - \frac{\varphi^2}{2} } \Omega\ts{err}~, \\
\Omega_\bot &= \Omega\ts{err} \sin \varphi \approx \varphi \Omega\ts{err}~, \\
\Omega &\approx \Omega_L + \prn{ 1 - \frac{\varphi^2}{2} } \Omega\ts{err} + \frac{\varphi^2\Omega\ts{err}^2}{2\Omega_L}~.
\label{Eq:parallel-shift}
\end{align}
While this geometry offers no suppression $\Omega\ts{err}$, the systematic uncertainty in the value of $\Omega\ts{err}$ due to apparatus misalignment is quadratically suppressed.  If $\Omega\ts{err}$ is independently measured with sufficient accuracy, it can be subtracted from the data.

The various systematic effects considered in this section are summarized in Table~\ref{Table:SystematicEffects}, which lists their estimated contribution to $\Delta \sR$.  The most significant estimated source of systematic uncertainty in our experiment is the effect of light shifts due to the residual ellipticity of the nominally linearly polarized probe beam.

\begin{table}
\caption{Estimated contribution of various systematic errors to $\Delta \sR$ for $|\mb{B}| = 7.1443~{\rm mG}$.  The atomic shot-noise-limited sensitivity of the setup under our experimental conditions, $N \approx 10^{12}~{\rm atoms}$ and $\Gamma\ts{rel} \approx 2\pi \times 1~{\rm Hz}$, is also listed for comparison, along with our anticipated experimental sensitivity (corresponding to $\delta \Omega = 2\pi \times 10~{\rm \mu Hz}$ in 1~second of integration).  An integration time of $10^6$~seconds is assumed. For $k=1$, $\chi_p = \hbar/c$, leading to a spin-gravity signal at the level $\Delta \sR \approx 3 \times 10^{-13}$.}
\medskip \begin{tabular}{ll} \hline \hline
\rule{0ex}{3.6ex} Description & $\Delta \sR$ \\
\hline
\rule{0ex}{3.6ex} Atomic shot-noise limit~~~~~~~~~~ & $4 \times 10^{-14}$ \\
\rule{0ex}{3.6ex} Anticipated sensitivity & $1 \times 10^{-12}$ \\
\rule{0ex}{3.6ex} Gyro-compass effect & $2 \times 10^{-14}$ \\
\rule{0ex}{3.6ex} Nuclear magnetic moments & $< 10^{-16}$ \\
\rule{0ex}{3.6ex} Nonlinear Zeeman effect & negligible \\
\rule{0ex}{3.6ex} Light shifts & $< 10^{-12}$ \\
\rule{0ex}{3.6ex} Spin-exchange collisions & $2 \times 10^{-18}$ \\
\rule{0ex}{3.6ex} Magnetic field gradients \& geometric phase~~~~ & $< 10^{-16}$ \\
\rule{0ex}{3.6ex} Wall collisions & $< 10^{-16}$ \\
\hline \hline
\end{tabular}
\label{Table:SystematicEffects}
\end{table}

\subsection{Gyro-compass effect}
\label{Sec:gyro-compass}

Because the experimental apparatus is attached to the Earth, while the atomic spins are decoupled from Earth's rotation, the experimental signal is sensitive to the rotation rate of the Earth, $\Omega_E/(2\pi) \approx 11.6~{\rm \mu Hz}$.  This effect, known as the gyro-compass effect \cite{Hec08} or the spin-rotation effect \cite{Ni10}, can be understood as the result of viewing an inertial system, the atomic spins, from a noninertial frame, the surface of the rotating Earth. Uncertainty in the magnitude of this $B$-odd systematic effect can be made quadratic in the misalignment of the experimental apparatus (Eq.~\ref{Eq:parallel-shift}) by orienting $\mb{B}$ along the axis of Earth's rotation $\hat{\bs{\Omega}}_E$ \cite{Ven92}.  This approach has been implemented as shown in the picture at the top left of the experimental setup diagram (Fig.~\ref{Fig:expt-setup}).

Including the gyro-compass effect adds a $B$-odd spin-precession frequency ($\Omega_o=\Omega_E \cos\theta$) to $\Omega_{85}$ and $\Omega_{87}$, where $\theta$ describes the misalignment between Earth's rotation axis $\hat{\bs \Omega}_E$ and $\mb{B}$.  Based on Eq.~\eqref{Eq:B-odd-shift}:
\begin{align}
\Delta \sR \approx \prn{\frac{\gamma_{87}-\gamma_{85}}{\gamma_{87}+\gamma_{85}}} \sbrk{ 4.12 \prn{\frac{\chi_p g \cos\phi}{\mu_0 B}}  - 2.4 \prn{\frac{\Omega_E \cos\theta}{\mu_0 B}} }~.
\end{align}
The angle $\phi$ is now the resultant angle between $\bs{g}$ and $\hat{\bs{\Omega}}_E$ ($\phi$ equals $90^\circ$ plus the latitude of the laboratory location, about $37^\circ$, so $\cos\phi \approx -0.6$).  We can control the orientation of $\mb{B}$ with respect to an auxiliary laser beam propagating along $\mb{\hat{z}}$ to a level of better than $10^{-5}$ \cite{Pus06-tilt}.  The long lever arm of the laser beam (in combination with GPS and aerial surveying) enables alignment of the auxiliary laser beam propagation direction with $\hat{\bs{\Omega}}_E$ to within $\approx 0.3^\circ \approx 5 \times 10^{-3}~{\rm rad}$, so that systematic uncertainty in the gyro-compass effect due apparatus misalignment is at the $3 \times 10^{-10}~{\rm Hz}$ level.  Thus errors due to the Earth's rotation can be well-controlled at our proposed level of sensitivity.

\subsection{Nuclear magnetic moments}
\label{Sec:nuclear-magnetic-moments}

Although nuclear magnetic moments are a thousand times smaller than $\mu_0$, their effect on the observed spin precession frequencies is clearly evident in the sample data shown in Fig.~\ref{Fig:GDM-data}, giving rise to a slow beating visible in the time-domain signal shown in the upper plot.  The nuclear magnetic moment modifies the Land\'e factors for the alkali ground state hyperfine levels \cite{Ale93,Auz10}:
\begin{align}
g_{F=I+\frac{1}{2}} & = \frac{2}{2I+1} - g_I \frac{\mu_N}{\mu_0} \frac{2I}{2I+1}~,  \label{Eq:Lande-g-factor-nucl-mag-mom-1} \\
g_{F=I-\frac{1}{2}} & = -\frac{2}{2I+1} - g_I \frac{\mu_N}{\mu_0} \frac{2(I+1)}{2I+1}~,
\label{Eq:Lande-g-factor-nucl-mag-mom-2}
\end{align}
where $g_I$ is the nuclear Land\'e factor ($g_I \approx 0.539$ for $^{85}$Rb, $g_I \approx 1.827$ for $^{87}$Rb \cite{Ari77}) $\mu_N$ is the nuclear magneton, and $\mu_N/\mu_0 \approx 5 \times 10^{-4}$.  This creates a difference in the Larmor frequencies for atoms in the two different ground state hyperfine levels
\begin{align}
\Delta \Omega\ts{nucl} &= \Omega_L \prn{F=I+\frac{1}{2}} - \Omega_L \prn{F=I-\frac{1}{2}} \nonumber \\
&= -2 g_I \mu_N B~.
\end{align}
For $B \approx 7.1443~{\rm mG}$ as in the data shown in Fig.~\ref{Fig:GDM-data}, $^{85}$Rb has $\Delta \Omega\ts{nucl} \approx -2\pi \times 5.9~{\rm Hz}$ and $^{87}$Rb has $\Delta \Omega\ts{nucl} \approx -2\pi \times 19.9~{\rm Hz}$.  Off-resonant synchronous optical pumping for the $^{85}$Rb $F=2$ state is much more efficient than that for the $^{87}$Rb $F=1$ state because of the smaller $\Delta \Omega\ts{nucl}$.  The signal from the $^{87}$Rb $F=1$ state is additionally suppressed relative to the signal from the $^{85}$Rb $F=2$ state because the re-pump laser beam depletes the $^{87}$Rb $F=1$ state and the probe laser light is farther detuned from the $^{87}$Rb D2 $F=1 \rightarrow F'$ resonance than from the $^{85}$Rb D2 $F=2 \rightarrow F'$ resonance.  Consequently, only the dominant signals from the $^{85}$Rb $F=3$ and $^{87}$Rb $F=2$ states, along with a much smaller signal from the $^{85}$Rb $F=2$ state which leads to the slow beating observed in the time-domain signal in Fig.~\ref{Fig:GDM-data}, are easily detectable in the data.

In our data analysis, the three observable resonances in the Fourier transformed optical rotation data are fit directly, and modeling demonstrates that neglecting the resonance associated with the $^{87}$Rb $F=1$ state in our fitting routine does not affect our analysis at the desired level of accuracy.  Fortunately, any first-order systematic effect associated with the nuclear magnetic moments manifests as a $B$-even systematic that is suppressed by $\sim \delta B/B \sim 10^{-9}$ in the comagnetometer signal $\Delta \sR$ as discussed in Sec.~\ref{Sec:systematic-effects-general}.

\subsection{Nonlinear Zeeman effect}
\label{Sec:NLZ}

The magnetic field also mixes Zeeman sublevels in different ground state hyperfine levels, leading the Zeeman effect to acquire a nonlinear dependence on $B$. In our experiment, the nonlinear Zeeman effect manifests as a splitting of the Larmor resonances \cite{Aco06,Aco08,Jen09,Pus11}.  The splitting of the resonances is symmetric about the unperturbed Larmor frequency and smaller than the linewidth, therefore, to leading order, it does not contribute any systematic shift to the spin precession frequencies.  The energy $E(F,M_F)$ of a particular ground state Zeeman sublevel ($M_F$ is the projection of $F$ along $\mb{\hat{z}}$) of an alkali atom is described by the Breit-Rabi formula \cite{Sob92}:
\begin{equation}
\begin{split}
E(F &= I \pm 1/2, M_F) = \\ -\frac{\sA\ts{hfs}}{4} &- g_I \mu_N B M_F \pm \frac{\sA\ts{hfs}}{4}(2I+1)\sqrt{ 1 + \frac{4M_F u}{2I+1} +u^2 }~,
\label{Eq:Breit-Rabi}
\end{split}
\end{equation}
where $\sA\ts{hfs}$ is the alkali atom's hyperfine structure constant and $u$ is the perturbation parameter given by:
\begin{align}
u \equiv \frac{g_J \mu_0 + g_I \mu_N}{2I+1}\frac{2B}{\sA\ts{hfs}} \approx \frac{4}{2I+1} \frac{\mu_0 B}{\sA\ts{hfs}}~,
\end{align}
where $g_J \approx 2$ is the Land\'e $g$-factor for the electron.  The Breit-Rabi formula (Eq.~\ref{Eq:Breit-Rabi}) can be expanded to second order in $u$ and the terms proportional to $u^2$ can be identified as the nonlinear Zeeman shifts $E\ts{nlz}$:
\begin{equation}
\begin{split}
E\ts{nlz} (F = &I \pm 1/2, M_F) = \\ & \pm u^2\frac{\sA\ts{hfs}}{8} \prn{ 2I + 1 } \prn{ 1 - \frac{4 M_F^2}{(2I+1)^2} }~.
\label{Eq:NLZ-energy}
\end{split}
\end{equation}
The term in Eq.~\eqref{Eq:NLZ-energy} proportional to $M_F^2$,
\begin{align}
\approx \mp \frac{8}{\prn{2I+1}^3}\frac{\mu_0^2 B^2}{\sA\ts{hfs}}M_F^2~,
\end{align}
causes a nonlinear Zeeman shift of the Larmor frequencies that splits a single spin-precession resonance into multiple resonances.  In the following we consider only the $F=I+1/2$ ground state hyperfine levels and define the unperturbed Larmor frequency $\Omega_L^{(0)}$ as the term linear in $B$,
\begin{align}
\Omega_L^{(0)} = \prn{ \frac{2}{2I+1}\mu_0 - g_I \mu_N \frac{2I}{2I+1} } B~.
\end{align}
For the $^{85}$Rb $F=3$ state, there appear six resonance frequencies split symmetrically about $\Omega_L^{(0)}$:
\begin{align}
\Omega_L^{(0)} \pm 5 \frac{8}{\prn{2I+1}^3}\frac{\mu_0^2 B^2}{\sA\ts{hfs}}~, \nonumber \\
\Omega_L^{(0)} \pm 3 \frac{8}{\prn{2I+1}^3}\frac{\mu_0^2 B^2}{\sA\ts{hfs}}~, \nonumber \\
\Omega_L^{(0)} \pm 1 \frac{8}{\prn{2I+1}^3}\frac{\mu_0^2 B^2}{\sA\ts{hfs}}~. \nonumber
\end{align}
(For the $^{87}$Rb $F=2$ state there are four resonance frequencies described by the latter four cases above.)  For $B \approx 7.1443~{\rm mG}$, as in the data shown in Fig.~\ref{Fig:GDM-data}, the maximum splitting of the resonance frequencies is $\approx 0.2~{\rm Hz}$ for the $^{85}$Rb $F=3$ state and $\approx 0.08~{\rm Hz}$ for the $^{87}$Rb $F=2$ state, in both cases smaller than the resonance linewidth of $\approx 1~{\rm Hz}$.  Any imbalance in the population of the Zeeman sublevels associated with the different resonances constitutes longitudinal spin polarization that does not contribute to the spin precession signal. Therefore, in some sense, the signal amplitudes for different resonances are naturally balanced. Thus the only apparent consequence of the nonlinear Zeeman effect under our experimental conditions is a slight broadening of the spin precession resonances.  Nonetheless, measurements will be carried out at different magnetic fields to test for any magnetic-field-dependent systematic errors.

\subsection{Light shifts}
\label{Sec:lightshifts}

The ac Stark effect due to the optical electric field of the probe beam can cause light shifts of Zeeman sublevels, leading to shifts of the measured precession frequencies for $^{85}$Rb and $^{87}$Rb.  In general, ac Stark shifts can be described in terms of scalar, vector, and tensor polarizabilities \cite{Auz10,Sta06}.  The scalar and tensor polarizabilities are described by rank-zero and rank-two operators, and thus their effect on atoms can be modeled as a fictitious static electric field along the light polarization axis;  the vector polarizability is described by a rank-one operator, and thus can be modeled as a fictitious static magnetic field along the light propagation direction $\mb{\hat{k}}$ \cite{Hap67,Mat68,Coh72,Rom99,Par02}.

For measurement of $\Omega_{85}$ and $\Omega_{87}$, it is the vector light shift in particular that causes the most significant systematic effect.  Although the probe beam is nominally linearly polarized and detuned far from the Doppler-broadened optical resonances, vector light shifts can still arise due to residual ellipticity $\epsilon$ induced in the beam due to birefringence of the vapor cell walls.  Measurements of the probe beam polarization before and after the cell using a ThorLABs PAX720IR1-T polarimeter system show that $\epsilon$ can be made $\lesssim 0.01^\circ \approx 2 \times 10^{-4}~{\rm rad}$.

The frequency shift $\Omega\ts{ac}$ associated with the vector polarizability is a $B$-odd systematic effect, and thus is not suppressed in the comagnetometer signal $\Delta \sR$. However, because $\mb{k}$ is orthogonal to $\mb{B}$, there is a geometric suppression according to Eq.~\eqref{Eq:orthogonal-shift}.  The quadratic correction term in Eq.~\eqref{Eq:orthogonal-shift} can be neglected in our case, and we have:
\begin{align}
\Omega\ts{ac} \approx \varphi \sin(2\epsilon) \Delta E\ts{ac}/\hbar \approx 2 \varphi \epsilon \Delta E\ts{ac}/\hbar~,
\label{Eq:acStark-estimate}
\end{align}
where $\Delta E\ts{ac}$ is the vector light shift between adjacent Zeeman sublevels ($\Delta M_F = 1$) for left-circularly polarized light along the quantization axis.  $\Delta E\ts{ac}$ can be estimated, for example, based on the formula from Ref.~\cite{Par02}:
\begin{align}
\Delta E\ts{ac} \approx - \frac{|\bra{ 5 S_{1/2} } er \ket{ 5 P_{1/2} }|^2}{9 \Delta\omega_{3/2}} g_F \abrk{ |\sE_0|^2 }~,
\end{align}
where $\bra{ 5 S_{1/2} } er \ket{ 5 P_{1/2} } \approx 3ea_0$ is the transition dipole matrix element between the $5 S_{1/2}$ and $5 P_{1/2}$ states, $a_0$ is the Bohr radius, $\Delta\omega_{3/2}$ is the detuning of the probe beam from the D2 resonance, $g_F$ is the ground state Land\'e factor, and $\abrk{ |\sE_0|^2 }$ is the average square of the optical electric field experienced by the atoms.  In calculating $\abrk{ |\sE_0|^2 }$, one must take into account the fact that the effective optical electric field experienced by the atoms is diluted by the ratio of the volume within the cell illuminated by the probe light beam to the total volume of the cell (for our experiment, the ratio $\approx 2 \times 10^{-3}$) since the atoms spend only a small fraction of their time in the probe light during the precession time \cite{Jen09}.  For our typical probe light power of $200~{\rm \mu W}$ and detuning of $\approx 3~{\rm GHz}$ below the center frequency of the Doppler-broadened $^{87}$Rb D2 $F=2 \rightarrow F'$ resonance, we estimate that the vector light shifts for $^{85}$Rb and $^{87}$Rb are given, respectively, by
\begin{align}
\Delta E\ts{ac}(85) &\approx -2\pi\hbar \times (3~{\rm Hz})~, \\
\Delta E\ts{ac}(87) &\approx -2\pi\hbar \times (1.3~{\rm Hz})~.
\end{align}
Because sensitive nonlinear magneto-optical effects can be used to directly measure the angle between $\mb{k}$ and $\mb{B}$ \cite{Pus06-tilt}, it is actually feasible in our setup to constrain $\varphi \lesssim 10^{-5}~{\rm rad}$.  Therefore, based on Eq.~\eqref{Eq:acStark-estimate}, $\Omega\ts{ac} \lesssim 10^{-8}~{\rm Hz}$, and consequently light shifts are not expected to prevent the experiment from reaching its sensitivity target.  Nevertheless, data will be taken at different probe light powers to check for any systematic effects related to light shifts.

\subsection{Spin-exchange collisions}
\label{Sec:SE-collisions}

The dual-isotope Rb comagnetometer relies on independent measurements of $\Omega_{85}$ and $\Omega_{87}$, so coupling between the two isotopes through spin-exchange (SE) collisions can produce a systematic error.  However, since the experiment is carried out in a bias field of $|\mb{B}| \sim 10~{\rm mG}$ and $\Omega_{85} \neq \Omega_{87}$, in the frame rotating with each isotope's precession frequency, the spin-polarization of the other isotope is time-averaged to nearly zero \cite{Har70}.  Nonetheless, there still appears a small SE frequency shift \cite{Hap73}.  Spin-exchange collisions tend to pull the precession frequencies toward a weighted average: SE collisions that transfer atoms between ground state hyperfine levels of a single isotope reduce the measured spin precession frequency since the gyromagnetic ratios have opposite signs; SE collisions between Rb isotopes shift $\Omega_{85}$ to a higher frequency and $\Omega_{87}$ to a lower frequency (cross-isotope SE shifts do not cancel because of the larger statistical weights of the $F=I+1/2$ hyperfine levels).

An estimate of the scale of the SE frequency shift $\Omega\ts{se}$ can be obtained by considering SE collisions between ground state hyperfine levels of each individual isotope.  Under our experimental conditions, where the SE collision rate $\gamma\ts{se} \ll \Omega_L$, we have Ref.~\cite{Hap73}:
\begin{align}
\Omega\ts{se} \approx - \frac{\gamma\ts{se}^2}{18 \Omega_L}\prn{1-\frac{1}{(2I+1)^2}}\prn{1-\frac{4}{(2I+1)^2}}~.
\end{align}
Under the experimental conditions for the data shown in Fig.~\ref{Fig:GDM-data} ($\gamma\ts{se} \approx 2\pi \times 1.3~{\rm Hz}$, $\Omega\ts{85} \approx 2\pi \times 3334~{\rm Hz}$, and $\Omega\ts{87} \approx 2\pi \times 5001~{\rm Hz}$):
\begin{align}
\Omega\ts{se}(85) \approx - 2\pi \times 2.3 \times 10^{-5}~{\rm Hz}~, \\
\Omega\ts{se}(87) \approx - 2\pi \times 1.3 \times 10^{-5}~{\rm Hz}~.
\end{align}
Crucially, SE frequency shifts are $B$-even and so their effect on the comagnetometer signal $\Delta \sR$ is described by Eq.~\eqref{Eq:B-even-shift}, thus suppressing any SE collision-related systematic effects by $\sim \delta B/B \sim 10^{-9}$.  Because of the suppression of $\Omega\ts{se}$ in $\Delta \sR$, systematic effects due to SE collisions between atoms are negligible in our experiment.

\subsection{Other systematic effects}
\label{Sec:other-systematics}

Another concern is the effect of magnetic field gradients which cause $^{85}$Rb and $^{87}$Rb atoms to, on a random basis, sample different magnetic fields, reducing the effectiveness of the comagnetometry scheme.  Field gradients are nulled using auxiliary measurements to $\lesssim 10^{-7}~{\rm G/cm}$ in all directions \cite{Pus06-grad}.  Effects of gradients are further reduced due to motional averaging in the evacuated antirelaxation-coated cells: atoms typically bounce off of the cell walls $\gtrsim 10^5$ times between interactions with the laser beam \cite{Bal10}.  For a sample of $\sim 10^{12}$~atoms, this creates uncertainty at the ${\rm nHz/\sqrt{Hz}}$ level, well below our statistical sensitivity to spin precession.  Furthermore, systematic frequency shifts related to the geometric (Berry's) phase \cite{Ber84} are proportional to gradients, and are estimated to be less than a nHz under typical experimental conditions based on the analysis of Ref.~\cite{Pen04}.

Wall collisions can produce quadrupolar splittings of spin precession frequencies due interaction of atomic spins with surface electric field gradients \cite{Ven92}.  In our experiment, wall collisions should produce negligibly small shifts of $\Omega_{85}$ and $\Omega_{87}$ since the vapor cell employs an amorphous antirelaxation coating and is spherical in shape, so that, to a high precision, there is no preferred direction in the cell.  We can estimate that in the worst-case scenario the contribution to a cell-related shift is on the order of the wall relaxation rate ($\sim 10^{-2}~{\rm Hz}$) times the square of the ratio of the size of the opening to the stem that contains the alkali metal sample ($\approx 10^{-2}~{\rm cm^2}$) to the inner surface area of the cell ($\approx 80~{\rm cm^2}$): $\lesssim 10^{-10}~{\rm Hz}$.

\section{Conclusion}
\label{Sec:conclusion}

An experiment measuring spin precession frequencies of overlapping ensembles of $^{85}$Rb and $^{87}$Rb atoms contained within an evacuated, antirelaxation-coated vapor cell can be used to search for presently unconstrained anomalous long-range spin-mass couplings.  Synchronous optical pumping with circularly polarized light is used to generate spin polarization transverse to an applied magnetic field and optical rotation of a linearly polarized probe beam is used to measure the $^{85}$Rb and $^{87}$Rb spin precession frequencies. The Earth is used as the source mass.  The present statistical sensitivity of the apparatus to spin precession frequencies is $10^{-4}~{\rm Hz}$ in one second of integration, with a shot-noise-projected sensitivity exceeding this level by over two orders of magnitude.  A variety of systematic errors are considered, and all known sources of error can be controlled at the $10^{-8}~{\rm Hz}$ level.

There are several promising and potentially more sensitive approaches to searching for long-range spin-mass couplings, including the use of spin-exchange-relaxation free (SERF) comagnetometers \cite{Kor02,Bro10,Smi11}, $^3$He/$^{129}$Xe free-precession comagnetometers \cite{Gem10}, and liquid state nuclear-spin comagnetometers \cite{Led12}.  However, there are experimental challenges to applying each of these alternative approaches to a search for long-range spin-mass couplings.  For example, it is potentially difficult to distinguish the coupling of spins to the local gravitational field from other lab-fixed backgrounds with a SERF comagnetometer \cite{Kor02}, and magnetic field gradients may be an issue for liquid state nuclear-spin comagnetometers \cite{Led12}.

These same techniques can also be applied to search for long-range anomalous spin-spin interactions using polarized electrons in the Earth \cite{Hun13}.

\acknowledgments

The authors are grateful to Dmitry Budker, Brian Patton, Szymon Pustelny, and Micah Ledbetter for invaluable discussions.  Important early contributions to the experiment were made by Khoa Nguyen, L. Rene Jacome, Eric Bahr, Srikanth Guttikonda, Delyana Delcheva, and Lok Fai Chan. We are indebted to Mohammad Ali for technical work on several parts of the apparatus.  This work was supported by the National Science Foundation under grants PHY-0652824 and PHY-0969666.  Any opinions, findings and conclusions or recommendations expressed in this material are those of the authors and do not necessarily reflect those of the National Science Foundation.

\end{document}